\documentclass[conference]{IEEEtran}
\IEEEoverridecommandlockouts
\usepackage{cite}
\usepackage{amsmath,amssymb,amsfonts}
\usepackage{algorithmic}
\usepackage{graphicx}
\usepackage{textcomp}
\usepackage{xcolor}
\usepackage{algorithm}
\usepackage{tabularx}
\usepackage{booktabs} 
\usepackage{subcaption}
\usepackage{multirow}
\usepackage{flushend}
\usepackage{tikz}
\def\BibTeX{{\rm B\kern-.05em{\sc i\kern-.025em b}\kern-.08em
    T\kern-.1667em\lower.7ex\hbox{E}\kern-.125emX}}

\newcommand\submittedtext{%
  \footnotesize This work has been submitted to the IEEE for possible publication. Copyright may be transferred without notice, after which this version may no longer be accessible.}
\newcommand\submittednotice{%
\begin{tikzpicture}[remember picture,overlay]
\node[anchor=north,yshift=-10pt] at (current page.north) {\fbox{\parbox{\dimexpr0.65\textwidth-\fboxsep-\fboxrule\relax}{\submittedtext}}};
\end{tikzpicture}%
}

\begin{document}

\bstctlcite{IEEEexample:BSTcontrol}

\title{Vision-Aided Near-Field Channel Estimation \\ for XL-MIMO}

\author{\IEEEauthorblockN{Sachira Karunasena\IEEEauthorrefmark{1},  Angel Lozano\IEEEauthorrefmark{3}, Tom Drummond\IEEEauthorrefmark{2}, and Rajitha Senanayake\IEEEauthorrefmark{1}}
\IEEEauthorblockA{\IEEEauthorrefmark{1}Department of Electrical and Electronic Engineering, University of Melbourne, Victoria, Australia. \\ \IEEEauthorrefmark{2}School of Computing and Information Systems, University of Melbourne, Victoria, Australia \\ 
\IEEEauthorrefmark{3}Department of Engineering, Universitat Pompeu Fabra, Barcelona, Spain\\
Email: \IEEEauthorrefmark{1}\IEEEauthorrefmark{2}\{s.karunasena, tom.drummond, rajitha.senanayake\}@unimelb.edu.au, \IEEEauthorrefmark{3}angel.lozano@upf.edu} }

\maketitle

\renewcommand\fbox{\fcolorbox{red}{white}}
\setlength{\fboxrule}{2pt} % Set fbox rule width to 2pt
\submittednotice

\begin{abstract}
In wideband extremely large-scale MIMO (XL-MIMO) systems, the transmitter (TX) is often in the near field of the base station array. The resulting spherical wavefront makes the line-of-sight channel depend on both the TX angle and distance. Existing polar-domain estimators sample both parameters across the entire coverage region, requiring many candidate positions and pilots. This paper proposes a vision-aided framework that uses the object carrying the TX, as seen in a camera image, to restrict the candidate positions to a small grid on that object. An on-grid vision-aided coarse localization (V-CL) algorithm obtains a coarse estimate from the grid points that best fit the received pilots, and an off-grid vision-aided Nelder-Mead refinement (V-NMR) algorithm refines it over continuous angle and distance, from which the channel is reconstructed. Simulations show that the proposed algorithms achieve higher accuracy than both wideband and narrowband polar-domain estimators while requiring far fewer pilots and candidate positions.
\end{abstract}


\section{Introduction}
\label{sec:intro}

Extremely large multiple-input multiple-output (XL-MIMO) is poised to be a central feature of 6G wireless networks \cite{andrews20246g}.
Owing to the extremely large number of antennas, XL-MIMO channel estimation might entail a prohibitively high
pilot overhead
\cite{cui2022channel}.
In addition, because of that large number of antennas and the relatively high frequencies at which 6G will operate, many (possibly most) users will operate in near-field conditions \cite{9422343}.

In the far field, methods based on compressive sensing can reduce the pilot overhead by exploiting angular-domain sparsity %through a discrete Fourier dictionary
\cite{lee2016channel, gao2016channel}. Specifically, the orthogonal matching pursuit (OMP) algorithm has been applied to narrowband channels \cite{lee2016channel} and the simultaneous OMP (SOMP) technique to wideband channels \cite{gao2016channel}. 

In the near field, however, the energy of each signal path spreads angularly,
destroying the sparsity \cite{cui2022channel}. A polar-domain SOMP (P-SOMP)
that considers both angles and distances is then more appropriate, along with off-grid refinements such as the polar-domain simultaneous iterative gridless weighting (P-SIGW) \cite{cui2022channel}. For wideband channels, bilinear pattern detection (BPD) shifts the dictionary index across subcarriers \cite{cui2023near}, the near-field beam-split-aware OMP (NBA-OMP) uses a separate dictionary per subcarrier \cite{elbir2023nba}, and the segmented P-SIGW (St-P-SIGW) applies P-SIGW separately to segments of subcarriers \cite{chen2024segmented}. Learning-based solutions are also emerging \cite{zhang2023nearml}. 
These polar-domain methods, whether traditional or learning-based, come at a steep computational cost. 
The dictionary must span every combination of angle and distance across the coverage region, whereby its size scales with the number of antennas times the number of sampled distances, and the ensuing grid is typically searched exhaustively.

If the base station (BS) had a coarse indication of the user's position, the search could be confined to a small region and most of the cost avoided. A camera at the BS can provide such an indication. Computer vision has indeed been proposed as a supplementary input for certain wireless communication tasks \cite{karunasena2025deep}. For channel estimation, though, the use of vision remains largely unexplored, despite the fact that images contain rich information about transmitter location, scatterers, and the propagation environment at large. 
Existing vision-aided channel estimation works target narrowband systems and rely on deep neural networks, trained on labeled image sequences, to predict the distance and angle of the user \cite{zhang2025vision}, leaving geometry-based, vision-aided channel estimation for near-field wideband channels as a promising, underexplored research avenue.

The present paper explores the conjunction of these avenues, considering near-field wideband channels with a single line-of-sight (LoS) path and a transmitter (TX) carried by a vehicle, although the formulation could be adapted to other objects or to persons. A single LoS path is representative of a roof-mounted TX with an unobstructed link to the BS.
It is posited that the object or person carrying the TX has already been identified among whatever objects are visible to the camera, and that its 3D bounding box and the camera calibration are available, meaning that the BS knows which object carries the TX, but not where on that object the TX sits. This is a deliberate premise, motivated by state-of-the-art results for TX identification in the far field \cite{karunasena2025deep}. Extending these results to the near field and to a multipath channel model are separate, highly relevant research problems, but outside the scope of the present work.
The contributions of this paper are:
\begin{itemize}
    \item A frequency-dependent (hence wideband) polar-domain dictionary built on a grid of candidate positions, sized by the TX object region rather than the number of antennas.
    \item An on-grid vision-aided coarse localization (V-CL) algorithm that selects the candidates on this grid that best match the received pilots, yielding coarse estimates of the TX angle and distance.
    \item An off-grid vision-aided Nelder-Mead refinement (V-NMR) algorithm that refines the V-CL output through a derivative-free search over the detected object containing the TX.
\end{itemize}

The proposed methods are shown to outperform both wideband and narrowband polar-domain benchmarks in normalized mean-square error (NMSE) and pilot overhead, at a lower or comparable runtime. They do so with two orders of magnitude fewer candidate positions and match the accuracy obtained with perfect knowledge of the TX position.

\section{System Model}
\label{sec:sys}

We consider a BS equipped with $N_{\mathrm{RF}}$ radio frequency (RF) chains and a uniform linear array (ULA) of $N_{\mathrm{bs}}$ antennas spaced by $d=\frac{\lambda_{\mathrm{c}}}{2}$, where $\lambda_{\mathrm{c}}=\frac{c}{f_{\mathrm{c}}}$ is the carrier wavelength, $f_{\mathrm{c}}$ is the carrier frequency, and $c$ is the speed of light. Antennas are indexed by $n = 1,\dots,N_{\mathrm{bs}}$ with positions $x_n = \delta_n d$, $\delta_n = n - (N_{\mathrm{bs}}+1)/2$. 
We consider a single-antenna TX transmitting from location $(\theta, r)$, where $r$ is the distance from the array center and $\theta = \sin\varphi \in [-1,1]$, with $\varphi$ being the azimuth angle relative to the array's normal.  The distance from the TX point to antenna $n$ is
\begin{equation}
r_n(\theta,r) = \sqrt{r^2 + \delta_n^2 d^2 - 2 r \theta \delta_n d}.
\label{eq:rn}
\end{equation}
The system features $N_{\mathrm{sc}}$ subcarriers spanning a bandwidth $B$, with subcarrier frequencies $f_m = f_{\mathrm{c}} + \left(m - \tfrac{N_{\mathrm{sc}}+1}{2}\right)\tfrac{B}{N_{\mathrm{sc}}}$ for $m = 1,\dots,N_{\mathrm{sc}}$.
The near-field steering vector at subcarrier $m$ is 
\begin{equation}
\boldsymbol{b}_m(\theta,r) = 
\left[ e^{-jk_mr_1},\, \ldots,\, e^{-jk_mr_{N_{\mathrm{bs}}}} \right]^{\mathsf{T}},
\label{eq:steering}
\end{equation}
where $k_m = \frac{2\pi f_m}{c}$, and the LoS channel coefficient between the TX point and antenna $n$ at subcarrier $m$ is then
\begin{equation}
h_{n,m}(\theta,r) = \frac{1}{2 k_m r_n} \left[ \boldsymbol{b}_m(\theta,r) \right]_n.
\label{eq:channel}
\end{equation}
Stacking the coefficients for the $N_{\mathrm{bs}}$ antennas yields the channel vector 
\begin{equation}
    \boldsymbol{h}_m(\theta,r) = \left[ h_{1,m}(\theta,r) \, \cdots \, h_{N_{\mathrm{bs}},m}(\theta,r) \right]^{\mathsf{T}} \in \mathbb{C}^{N_{\mathrm{bs}} \times 1},
\end{equation}
at subcarrier $m$.
Across all subcarriers, the channel matrix is 
\begin{equation}
    \boldsymbol{H}(\theta,r) = \left[ \boldsymbol{h}_1(\theta,r)\, \cdots\, \boldsymbol{h}_{N_{\mathrm{sc}}}(\theta,r) \right] \in \mathbb{C}^{N_{\mathrm{bs}}\times N_{\mathrm{sc}}}.
\end{equation}

The BS features a hybrid receiver. For each of $p = 1,\dots,P$ pilot symbols, the BS applies a random analog combining matrix 
$\boldsymbol{A}_p \in \{\pm 1/\sqrt{N_{\mathrm{bs}}}\}^{N_{\mathrm{RF}} \times N_{\mathrm{bs}}}$ to map the $N_{\mathrm{bs}}$ antennas onto the $N_{\mathrm{RF}}$ chains before digitization.
This random combiner acts as a compressed-sensing measurement matrix. 

The channel is block-faded, remaining unchanged during the pilot transmissions.
The received signal for pilot $p$ at subcarrier $m$ is, after the hybrid combiner,
\begin{equation}
\boldsymbol{y}_{p,m} = \boldsymbol{A}_p \boldsymbol{h}_m(\theta,r) + \boldsymbol{A}_p \boldsymbol{n}_{p,m},
\end{equation}
where $\boldsymbol{n}_{p,m} \sim \mathcal{N}_{\mathbb{C}}(\boldsymbol{0}, \sigma^2 \boldsymbol{I}_{N_{\mathrm{bs}}})$ with $\sigma^2$ the noise variance per antenna, independent across pilots and subcarriers. Stacking the $P$ observations we can write 
\begin{equation}
\boldsymbol{y}_m = \boldsymbol{A} \boldsymbol{h}_m(\theta,r) + \tilde{\boldsymbol{n}}_m,
\label{eqn:pilots}
\end{equation}
where
\begin{align}
&\boldsymbol{y}_m = [\boldsymbol{y}_{1,m}^{\mathsf{T}}, \dots, \boldsymbol{y}_{P,m}^{\mathsf{T}}]^{\mathsf{T}} \in \mathbb{C}^{PN_{\mathrm{RF}} \times 1}, \\
&\boldsymbol{A} = [\boldsymbol{A}_1^{\mathsf{T}}, \dots, \boldsymbol{A}_P^{\mathsf{T}}]^{\mathsf{T}} \in \mathbb{C}^{PN_{\mathrm{RF}} \times N_{\mathrm{bs}}}, \\
&\tilde{\boldsymbol{n}}_m = [(\boldsymbol{A}_1 \boldsymbol{n}_{1,m})^{\mathsf{T}}, \dots, (\boldsymbol{A}_P \boldsymbol{n}_{P,m})^{\mathsf{T}}]^{\mathsf{T}} \in \mathbb{C}^{PN_{\mathrm{RF}} \times 1} . \label{eq:noise}
\end{align}
The spatial covariance of the above noise is
\begin{equation}
\sigma^2\,\mathrm{blkdiag}\big(\boldsymbol{A}_1\boldsymbol{A}_1^*, \ldots, \boldsymbol{A}_P\boldsymbol{A}_P^*\big) = \sigma^2\boldsymbol{C},
\label{eq:noisecov}
\end{equation}
where $(\cdot)^*$ denotes the conjugate transpose, and, from the Cholesky factorization $\boldsymbol{C} = \boldsymbol{L}\boldsymbol{L}^*$, noise-whitened pilot observations can be obtained as
\begin{align}
\bar{\boldsymbol{y}}_m & = \boldsymbol{L}^{-1}\boldsymbol{y}_m \\
& = \boldsymbol{\psi}_m(\theta,r) + \bar{\boldsymbol{n}}_m,
\label{eq:whiten}
\end{align}
where $\boldsymbol{\psi}_m(\theta,r) = \boldsymbol{L}^{-1} \boldsymbol{A}\boldsymbol{h}_m(\theta,r)$ and $\bar{\boldsymbol{n}}_m $ is now white.
We then perform channel estimation based on the whitened observations $\bar{\boldsymbol y}_m$, together with the vision data.

\section{Problem Formulation}
\label{sec:problem_formul}

From \eqref{eq:channel},
the true channel can be related to the channel corresponding to an arbitrary candidate position $(\theta', r')$ via
\begin{align}
    h_{n,m}(\theta,r) & = \frac{r_n'}{r_n}\,e^{-jk_m(r_n-r_n')}\cdot \frac{1}{2 k_m r_n'}\, e^{-jk_mr_n'} \\
    & = \frac{r_n'}{r_n}\,e^{-jk_m(r_n-r_n')}\cdot h_{n,m}(\theta',r'),
    \label{eq:channel_with_candidate}
\end{align}
where $r_n' = r_n(\theta', r')$.
For candidates close to the TX, i.e., with an angular offset much smaller than the array beamwidth $2/N_{\mathrm{bs}}$ \cite{balanis2016antenna}, the formulation can be simplified by means of
$r_n-r_n' \approx r-r'$ and $\frac{r_n'}{r_n} \approx \frac{r'}{r}$ for all $n$. Then,
\begin{equation}
    h_{n,m}(\theta,r) \approx \frac{r'}{r}\,e^{-jk_m(r-r')}\cdot h_{n,m}(\theta',r') ,
    \label{eq:chn_approx}
\end{equation}
which, stacked over all the antennas, gives
\begin{equation}
\boldsymbol{h}_m(\theta,r) \approx \frac{r'}{r}\, e^{-jk_m(r-r')} \, \boldsymbol{h}_m(\theta',r').
\label{eq:chn_est}
\end{equation}
Plugging \eqref{eq:chn_est} into \eqref{eq:whiten} gives
\begin{equation}
\bar{\boldsymbol{y}}_m \approx \!\!\!\!\!\!\!\! \underbrace{ \frac{r'}{r}\, e^{-jk_m(r-r')}}_{\text{complex gain on subcarrier } m } \!\!\!\!\!\!\!\! \boldsymbol{\psi}_m(\theta',r') + \bar{\boldsymbol{n}}_m,
\label{eq:whiten_cand}
\end{equation}
where the unknown complex gain per subcarrier, $\alpha_m \in \mathbb{C}$, should be estimated from the pilots, jointly with the position. Altogether, the problem becomes
\begin{equation}
\min_{\theta',r'}\sum_{m=1}^{N_{\mathrm{sc}}}\min_{\alpha_m\in\mathbb C}\left\|\bar{\boldsymbol y}_m - \alpha_m\,\boldsymbol\psi_m(\theta',r')\right\|_2^2.
\label{eq:opt-old}
\end{equation}
For every candidate $(\theta',r')$, the inner minimization is a linear least-squares problem with solution
\begin{equation}
\begin{split}
 \operatorname*{arg\,min}_{\alpha_m\in\mathbb{C}} \left\| \bar{\boldsymbol{y}}_m - \alpha_m\, \boldsymbol{\psi}_m(\theta',r') \right\|_2^2 = \frac{\boldsymbol{\psi}_m^*(\theta',r')\,\bar{\boldsymbol{y}}_m} {\|\boldsymbol{\psi}_m(\theta',r')\|^2},
\end{split}
\label{eq:lsgain}
\end{equation}
which is the pilot-based estimate for $\frac{r'}{r}e^{-jk_m(r-r')}$, whereby \eqref{eq:opt-old} becomes
\begin{equation}
\min_{\theta',r'}\sum_{m=1}^{N_{\mathrm{sc}}}
\left\|\bar{\boldsymbol y}_m - \frac{\boldsymbol{\psi}_m^*(\theta',r')\,\bar{\boldsymbol{y}}_m} {\|\boldsymbol{\psi}_m(\theta',r')\|^2} \, \boldsymbol\psi_m(\theta',r')\right\|_2^2.
\label{eq:opt}
\end{equation}
Denoting by $(\hat\theta,\hat r)$ the solution of \eqref{eq:opt}, the channel estimate at subcarrier $m$, $\hat{\boldsymbol{h}}_m(\hat\theta,\hat r)$, is obtained by replacing the complex gain in \eqref{eq:chn_est} with its estimate \eqref{eq:lsgain} at that position, i.e.,
\begin{equation}
\hat{\boldsymbol{h}}_m(\hat\theta,\hat r) = \frac{\boldsymbol{\psi}_m^*(\hat\theta,\hat r)\,\bar{\boldsymbol{y}}_m}{\|\boldsymbol{\psi}_m(\hat\theta,\hat r)\|^2}\, \boldsymbol{h}_m(\hat\theta,\hat r).
\label{eq:recon}
\end{equation}
Stacking these estimates over all subcarriers yields the estimated channel matrix $\hat{\boldsymbol{H}}(\hat\theta,\hat r)$, whose accuracy is quantified by the NMSE,
\begin{equation}
\mathrm{NMSE} = \mathbb{E}\!\left(\frac{\|\boldsymbol{H}(\theta,r) - \hat{\boldsymbol{H}}(\hat\theta,\hat r)\|_F^2}{\|\boldsymbol{H}(\theta,r)\|_F^2}\right).
\label{eq:nmse}
\end{equation}
The approximation in \eqref{eq:chn_approx}--\eqref{eq:chn_est} is exact at the TX position and degrades for more distant candidates, whose channels can then no longer be matched by a single gain per subcarrier, which is what allows \eqref{eq:opt} to discriminate among candidates. Our goal is therefore to use the camera image to confine the search to candidates close to the TX, where the approximation holds, and to refine the resulting estimate off the grid.

\section{Dataset Generation}
\label{sec:sim}

All the TX positions considered in this work are in the radiative near field as determined by the Fraunhofer distance \cite{10070575}.
Evaluating the proposed methods requires each possible channel to be paired with an image in which the TX-carrying object is at the very position that would produce that channel. As, to the best of our knowledge, no public dataset provides such paired data, we combine real images with synthetic channels. The TX positions are derived from actual vehicles 
in a calibrated camera dataset, such that the angle-distance pairs $(\theta, r)$ follow realistic scene geometry; the corresponding near-field channels are generated from \eqref{eq:channel}.
DAIR-V2X is used, a vehicular infrastructure dataset for a fixed roadside camera \cite{yu2022dair}, with the TX being virtually positioned at a random point on the roof of a randomly chosen vehicle.

\begin{figure}
    \centering
    \includegraphics[clip, trim=4cm 2.5cm 0cm 0cm, width=\linewidth]{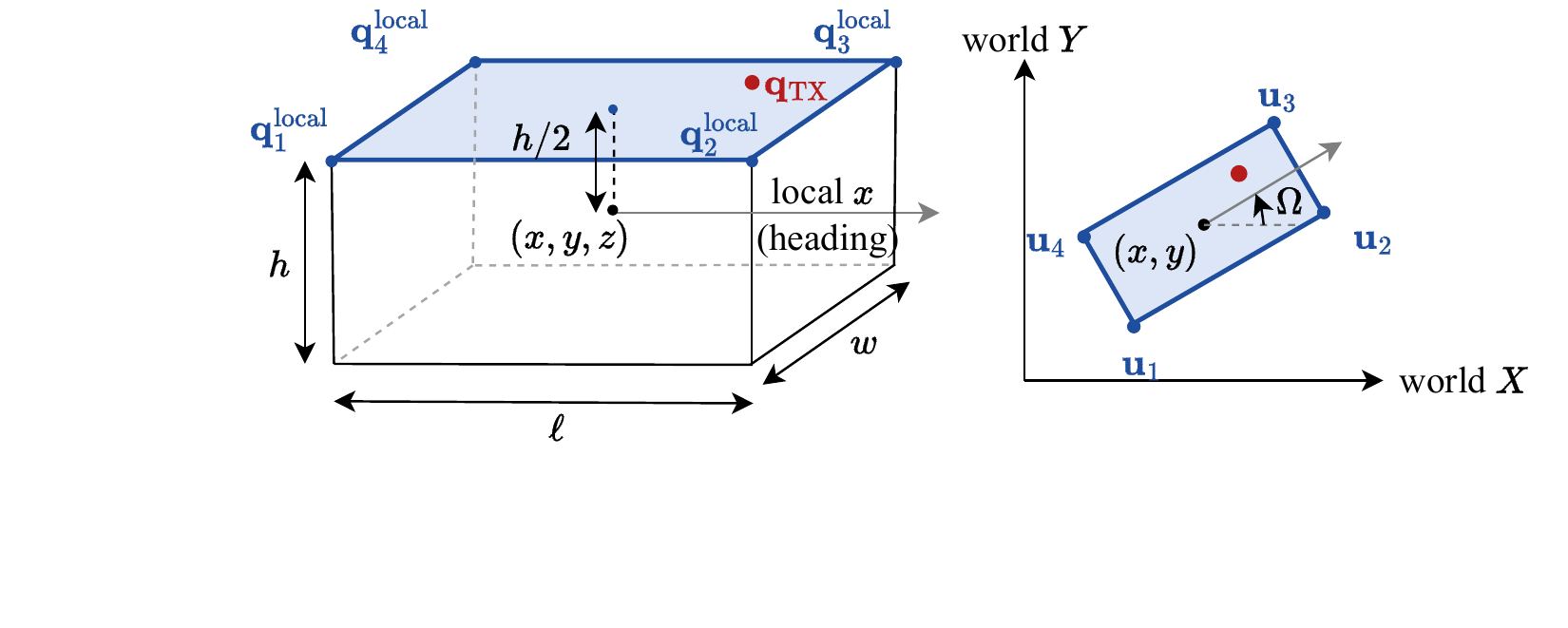}
    \caption{TX placement on the roof. Left: local frame, with the roof at $h/2$ above the center $(x,y,z)$. Right: top view in the world frame, rotated by $\Omega$ and translated to $(x,y)$. The TX (red) is at a random roof point $\mathbf{q}_{\mathrm{TX}}$.}
    \label{fig:tx_placement}
\end{figure}

Each vehicle is annotated with a 3D bounding box of center $(x, y, z)$, its length, width, and height $(\ell, w, h)$, and yaw $\Omega$. Its roof corners $\{\boldsymbol{u}_k\}_{k=1}^4$ are obtained by rotating the local roof rectangle $\{\boldsymbol{q}_k^{\mathrm{local}}\}_{k=1}^4$ by $\Omega$ and translating it to the vehicle's center, i.e.,
\begin{equation}
    \boldsymbol{u}_k = \boldsymbol{R}_{\Omega}\,\boldsymbol{q}_k^{\mathrm{local}} + [x, y, z]^{\mathsf{T}}, \quad k = 1,\ldots,4,
\label{eqn:roof-coord}
\end{equation}
given $\boldsymbol{q}_k^{\mathrm{local}} = [\pm \ell/2,\, \pm w/2,\, h/2]^{\mathsf{T}}$ as depicted in Fig. \ref{fig:tx_placement}. 
The TX is then placed at a uniformly random point $\boldsymbol{q}_{\mathrm{TX}}$ on this roof quadrilateral.
Such $\boldsymbol{q}_{\mathrm{TX}}$ is then converted to the ground-truth pair $(\theta, r)$ relative to the camera, which is co-located with the array center. First, its position is rotated into the camera frame,
\begin{equation}
\boldsymbol{v} = \boldsymbol{R}\,(\boldsymbol{q}_{\mathrm{TX}} - \boldsymbol{c}),
\label{eq:v}
\end{equation}
given the world-to-camera rotation $\boldsymbol{R}$ and the camera position $\boldsymbol{c}$. Then, $\varphi = \tan^{-1}(v_x, v_z)$, $\theta = \sin\varphi$, and $r$ is the horizontal distance between $\boldsymbol{q}_{\mathrm{TX}}$ and $\boldsymbol{c}$. As in \cite{cui2022channel, cui2023near}, only the azimuth is considered, with the array along the camera's $x$-axis and its broadside along the optical axis.

\section{Proposed Solution}
\label{sec:methodology}

\subsection{Vision-aided Grid Forming}
\label{sec:v-pd}

\begin{algorithm}[t]
\small
\caption{Vision-aided Grid Construction}
\label{alg:vcl-grid}
\begin{algorithmic}[1]
\raggedright
\REQUIRE Roof corners in the raw image $\{\boldsymbol{u}_k^{\mathrm{img}}\}_{k=1}^4$, roof height $z_0$; camera intrinsics $\boldsymbol{K}$, distortion $\boldsymbol{D}$, rotation $\boldsymbol{R}$, center $\boldsymbol{c}$; grid resolution $G$; canvas size $S$
\ENSURE Candidate set $\mathcal{C} = \{(\theta_i, r_i, \varphi_i)\}_{i=1}^{N}$, $N = G^2$

\STATE Compute homography $\boldsymbol{J}$: \\
$\{\boldsymbol{u}_k^{\mathrm{img}}\}_{k=1}^4 \to \{(0,0),(S,0),(S,S),(0,S)\}$
\FOR{$row = 0$ \TO $G-1$}
  \FOR{$col = 0$ \TO $G-1$}
     \STATE $i \leftarrow row \cdot G + col + 1$
     \STATE Rectified cell coordinate: $(u_{\mathrm r}, v_{\mathrm r}) = \big(col/(G-1)\cdot S,\ row/(G-1)\cdot S\big)$
     \STATE Map back to raw image: $(u_i, v_i) = \boldsymbol{J}^{-1}(u_{\mathrm r}, v_{\mathrm r})$
     \STATE Undistort $(u_i, v_i)$ with $\boldsymbol{K}, \boldsymbol{D}$ \cite{zhang2000flexible} to get $(\xi_i, \zeta_i)$
     \STATE Camera-frame ray: $\boldsymbol{d}_{\mathrm{cam}} = [\xi_i, \zeta_i, 1]^{\mathsf{T}}$
     \STATE Rotate to world frame: $\boldsymbol{d}_{\mathrm{world}} = \boldsymbol{R}^{\mathsf{T}}\boldsymbol{d}_{\mathrm{cam}}$
     \STATE Plane-intersection distance: $s = (z_0-c_z)/(\boldsymbol{d}_{\mathrm{world}})_z$
     \STATE World point: $\boldsymbol{p}_i = \boldsymbol{c} + s\,\boldsymbol{d}_{\mathrm{world}}$
     \STATE Angle and range: $\boldsymbol{v} = \boldsymbol{R}(\boldsymbol{p}_i - \boldsymbol{c})$, $\varphi_i = \tan^{-1}(v_x, v_z)$, \\
    $\theta_i = \sin\varphi_i$, $r_i$ = horizontal distance between $\boldsymbol{p}_i$ and $\boldsymbol{c}$
  \ENDFOR
\ENDFOR
\STATE $\mathcal{C} \leftarrow \{(\theta_i, r_i, \varphi_i)\}_{i=1}^{N}$
\RETURN $\mathcal{C}$
\end{algorithmic}
\end{algorithm}

In contrast with \cite{cui2022channel, cui2023near}, we intend to design
a grid concentrated on the identified TX rather than the entire field of view. 
The roof corners from \eqref{eqn:roof-coord} are projected into the raw image using the camera calibration \cite{zhang2000flexible}, and the resulting pixel corners $\{\boldsymbol{u}_k^{\mathrm{img}}\}_{k=1}^4$ are rectified into an $S \times S$ canvas, and a $G\times G$ grid is then overlaid. Each grid point is subsequently mapped to a point in the physical world using Alg.~\ref{alg:vcl-grid}.
This mapping first undistorts each pixel \cite{zhang2000flexible}, casts it as a ray into the world, and intersects the ray with the known roof-plane height to obtain a real-world point, which is converted into the angle and range used throughout the pipeline (steps~7--12 of Alg.~\ref{alg:vcl-grid}). Repeating this for every point yields the full candidate set $\mathcal{C}$, which feeds in as an input for Alg.~\ref{alg:vcl}.

Each grid point $(\theta_i, r_i)$ produced by Alg.~\ref{alg:vcl-grid} is purely spatial, and does not depend on frequency. 
For wideband channel estimation, each point is converted to the candidate channel $\boldsymbol{h}_m(\theta_i, r_i)$ at every subcarrier $m$. 
% Each spatial grid point yields a distinct candidate channel at every subcarrier.

\subsection{Vision-aided Coarse Localization}
\label{sec:v-cl}

\begin{algorithm}[t]
\small
\caption{Vision-Aided Coarse Localization (V-CL)}
\label{alg:vcl}
\begin{algorithmic}[1]
\REQUIRE Candidate set $\mathcal{C}=\{(\theta_i,r_i,\varphi_i)\}_{i=1}^{N}$ (Alg.~1); array offsets $\{\delta_n\}_{n=1}^{N_{\mathrm{bs}}}$, spacing $d$; wavenumbers $\{k_m\}_{m=1}^{N_{\mathrm{sc}}}$; combining matrix $\boldsymbol{A}$; whitening filter $\boldsymbol{L}$; whitened observations $\{\bar{\boldsymbol{y}}_m\}_{m=1}^{N_{\mathrm{sc}}}$; neighborhood margin $\eta$; confidence gate $\gamma$
\ENSURE Coarse estimate $(\hat{\theta}_{\mathrm{VCL}},\hat{r}_{\mathrm{VCL}})$
\FOR{$i=1$ \TO $N$}
  \STATE Antenna distances, $n=1,\ldots,N_{\mathrm{bs}}$: \\
  \hspace{\algorithmicindent}$r_{i,n}=r_n(\theta_i,r_i)=\sqrt{r_i^2+\delta_n^2d^2-2r_i\theta_i\delta_n d}$
  \FOR{$m=1$ \TO $N_{\mathrm{sc}}$}
    \STATE Channel at the candidate \eqref{eq:channel}: \\
    \hspace{\algorithmicindent}$[\boldsymbol{h}_m(\theta_i,r_i)]_n=\frac{1}{2k_m r_{i,n}}\,e^{-jk_m r_{i,n}}$
    \STATE Whitened atom: $\boldsymbol{\psi}_{i,m}=\boldsymbol{L}^{-1}\boldsymbol{A}\,\boldsymbol{h}_m(\theta_i,r_i)$
    \STATE Least-squares estimate \eqref{eq:lsgain}: $\hat{\alpha}_{i,m} = \dfrac{\boldsymbol{\psi}_{i,m}^* \bar{\boldsymbol{y}}_m}{\boldsymbol{\psi}_{i,m}^* \boldsymbol{\psi}_{i,m}}$
    \STATE Residual: $e_{i,m} = \|\bar{\boldsymbol{y}}_m - \hat{\alpha}_{i,m} \boldsymbol{\psi}_{i,m}\|_2^2$
  \ENDFOR
  \STATE Normalized cost: \\
  \hspace{\algorithmicindent}$J_i^{\mathrm{dB}}=10\log_{10} \dfrac{\sum_{m} e_{i,m}}{\sum_{m}\|\bar{\boldsymbol{y}}_m\|_2^2} $
\ENDFOR
\STATE Best candidate: $i^\star = \arg\min_i J_i^{\mathrm{dB}}$, $\tau = J_{i^\star}^{\mathrm{dB}} + \eta$
\STATE Neighborhood: $\mathcal{N}=\{i: J_i^{\mathrm{dB}}\le\tau\}$
\IF{$J_{i^\star}^{\mathrm{dB}}\le\gamma$}
  \STATE Confidence weights: $w_i = 10^{-(J_i^{\mathrm{dB}} - J_{i^\star}^{\mathrm{dB}})/10}$, $i \in \mathcal{N}$
 \STATE Weighted angle: \\$\hat\varphi_{\mathrm{VCL}} = \tan^{-1}\!\big(\sum_{i\in\mathcal{N}} w_i \sin\varphi_i,\ \sum_{i\in\mathcal{N}} w_i \cos\varphi_i\big)$
\ELSE
  \STATE $\hat{\varphi}_{\mathrm{VCL}}=\varphi_{i^\star}$
\ENDIF
\STATE $\hat{\theta}_{\mathrm{VCL}}=\sin\hat{\varphi}_{\mathrm{VCL}}$, \ $\hat{r}_{\mathrm{VCL}}=r_{i^\star}$
\RETURN $(\hat{\theta}_{\mathrm{VCL}},\hat{r}_{\mathrm{VCL}})$
\end{algorithmic}
\end{algorithm}

Alg.~\ref{alg:vcl} summarizes the proposed on-grid V-CL algorithm, which solves \eqref{eq:opt} with the search restricted to the vision-aided candidate set $\mathcal{C}$. For each $(\theta_i, r_i, \varphi_i) \in \mathcal{C}$, we evaluate $\boldsymbol{h}_m(\theta_i, r_i)$ via \eqref{eq:channel}
and pass it through the combining matrix and the whitening filter to obtain $\boldsymbol{\psi}_{i,m}$. As the complex gain is unknown, we compute the least-squares estimate \eqref{eq:lsgain} that best fits the whitened pilots at each candidate, and the residual energy left unexplained by that candidate; this embodies the inner minimization in \eqref{eq:opt}. The residual sum over all subcarriers gives the normalized cost of candidate $i$ and the best-fitting candidate $i^\star$ is the one with the smallest cost.

Rather than relying on the angle of $i^\star$ alone, we account for cases where several nearby candidates fit almost equally well. We collect the neighborhood $\mathcal{N}$ of candidates whose cost lies within a margin $\eta$ of that of $i^\star$ (steps~11--12). If the cost of $i^\star$ falls below a confidence gate $\gamma$, we take a confidence-weighted circular mean of the azimuths in $\mathcal{N}$ (steps~13--15); otherwise, we take the azimuth of $i^\star$ directly (steps~16--17). The coarse angle estimate $\hat{\theta}_{\mathrm{VCL}}$ follows from this azimuth, while the distance is taken directly from $i^\star$ without averaging (step~19), as averaging it was found to make little practical difference to the resulting channel estimate.

\subsection{Vision-aided Nelder-Mead Refinement}
\label{sec:v-nmr}

\begin{algorithm}[t]
\small
\caption{Vision-Aided Nelder-Mead Refinement (V-NMR)}
\label{alg:vnmr}
\begin{algorithmic}[1]
\raggedright
\REQUIRE Inputs of Algorithm~\ref{alg:vcl}; distance bound $r_{\max}$; iteration budget $N_{\mathrm{it}}$
\ENSURE Refined estimate $(\hat\theta_{\mathrm{VNMR}}, \hat r_{\mathrm{VNMR}})$
\item[] \textbf{Initialization stage}
\STATE Obtain the coarse estimate $(\hat\theta_{\mathrm{VCL}}, \hat r_{\mathrm{VCL}})$ by Algorithm~\ref{alg:vcl}
\item[] \textbf{Refinement stage}
\STATE Clipped distance at a candidate $(\theta',r')$: $\tilde r = \min(r', r_{\max})$
\STATE Cost: $J(\theta',r') = \dfrac{\sum_{m} e_m}{\sum_{m}\|\bar{\boldsymbol{y}}_m\|_2^2}$, \\
with $e_m$ computed as in steps 2--7 of Algorithm~\ref{alg:vcl} at $(\theta', \tilde r)$
\STATE Minimize $J$ by Nelder-Mead~\cite{nelder1965simplex}: \\
$(\theta^\star, r^\star) = \textsc{NelderMead}\big(J,\,(\hat\theta_{\mathrm{VCL}},\hat r_{\mathrm{VCL}}),\,N_{\mathrm{it}}\big)$
\STATE $\hat\theta_{\mathrm{VNMR}} = \theta^\star$, \ $\hat r_{\mathrm{VNMR}} = \min(r^\star, r_{\max})$
\RETURN $(\hat\theta_{\mathrm{VNMR}}, \hat r_{\mathrm{VNMR}})$
\end{algorithmic}
\end{algorithm}
Our proposed off-grid V-NMR algorithm, summarized in Alg.~\ref{alg:vnmr}, refines the coarse estimate into a continuous-valued one. It first retrieves $(\hat\theta_{\mathrm{VCL}}, \hat r_{\mathrm{VCL}})$ from Alg.~\ref{alg:vcl}, and then solves \eqref{eq:opt} off the grid, i.e., over  continuous candidates $(\theta', r')$. 
Since the search could otherwise drift to distances beyond the TX vehicle, where the cost becomes nearly insensitive to the distance, the candidate distance is clipped at $r_{\mathrm{max}}$, the distance from the camera to the farthest point of the vehicle's bounding box.
The cost $J(\theta',r')$ is then computed exactly as in V-CL
and minimized via the derivative-free Nelder-Mead simplex method~\cite{nelder1965simplex}, which requires only evaluations of the cost, initialized at the V-CL estimate, with the initial simplex obtained by perturbing its azimuth and distance by 5\%, and run for at most $N_{\mathrm{it}}$ iterations. The distance of the resulting minimizer is again clipped at $r_{\mathrm{max}}$, giving the refined estimate $(\hat\theta_{\mathrm{VNMR}}, \hat r_{\mathrm{VNMR}})$.

\section{Performance Evaluation}
\label{sec:results}

% \begin{table}[!t]
% \centering
% \small
% \caption{Parameters }
% \label{tab:sim-params}
% \begin{tabularx}{\linewidth}{Xc}
% \toprule
% Parameter & Value \\
% \midrule
% Number of BS antennas, $N_{\mathrm{bs}}$ & 256 \\
% Number of RF chains, $N_{\mathrm{RF}}$ & 4 \\
% Carrier frequency, $f_\text{c}$ & 15~GHz \\
% Bandwidth, $B$ & 100~MHz \\
% Number of subcarriers, $N_{\mathrm{sc}}$ & 256 \\
% Number of pilots, $P$ & 32 \\
% Vision grid resolution, $G\times G$ & 10 $\times$ 10  \\
% Canvas size, $S\times S$ & 300 $\times$ 300\\
% Number of distance bins & 10 \\
% Samples per distance bin & 50 \\
% Total near-field samples & 500 \\
% Nelder-Mead iteration budget, $N_{\mathrm{it}}$ & 25 \\
% Neighborhood margin, $\eta$ & 3~dB \\
% Confidence gate, $\gamma$ & $-6$~dB \\
% \bottomrule
% \end{tabularx}
% \end{table}

\begin{table}[t]
\caption{Parameters}
\label{tab:sim-params}
\centering
\footnotesize
\setlength{\tabcolsep}{0pt}
\begin{tabular*}{\columnwidth}{@{\extracolsep{\fill}}lrlr@{}}
\toprule
Antennas, $N_{\mathrm{bs}}$ & 256 & Pilots, $P$ & 32 \\
RF chains, $N_{\mathrm{RF}}$ & 4 & Grid, $G \times G$ & $10 \times 10$ \\
Carrier, $f_{\mathrm{c}}$ & 15~GHz & Canvas, $S \times S$ & $300 \times 300$ \\
Bandwidth, $B$ & 100~MHz & NM iterations, $N_{\mathrm{it}}$ & 25 \\
Subcarriers, $N_{\mathrm{sc}}$ & 256 & Margin $\eta$, gate $\gamma$ & 3, $-6$~dB \\
\bottomrule
\end{tabular*}
\end{table}

\begin{figure*}[!t]
    \centering
    \begin{subfigure}[b]{0.49\textwidth}
        \centering
        \includegraphics[width=\linewidth]{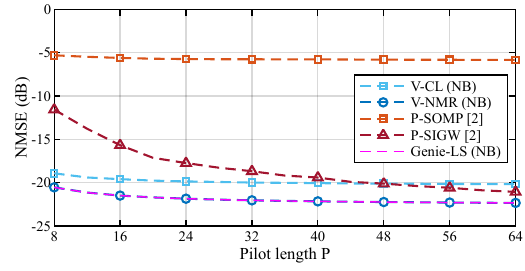}
        \caption{NMSE versus pilot length $P$ (SNR = 10~dB)}
        \label{fig:nb_pilot}
    \end{subfigure}
    \hfill
    \begin{subfigure}[b]{0.49\textwidth}
        \centering
        \includegraphics[width=\linewidth]{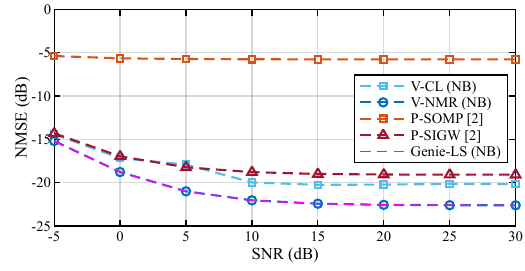}
        \caption{NMSE versus received SNR ($P = 32$)}
        \label{fig:nb_snr}
    \end{subfigure}
    \caption{Comparison under the narrowband model.}
    \label{fig:nb}
\end{figure*}

\begin{figure*}[!t]
    \centering
    \begin{subfigure}[b]{0.49\textwidth}
        \centering
        \includegraphics[width=\linewidth]{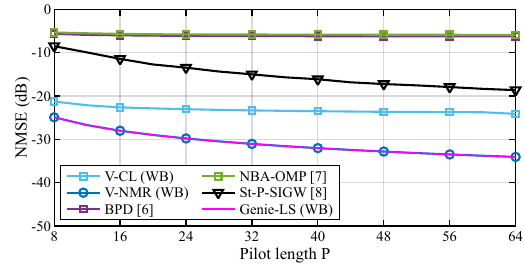}
        \caption{NMSE versus pilot length $P$ (SNR = 10~dB)}
        \label{fig:wb_pilot}
    \end{subfigure}
    \hfill
    \begin{subfigure}[b]{0.49\textwidth}
        \centering
        \includegraphics[width=\linewidth]{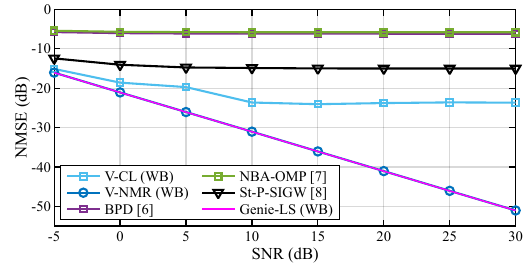}
        \caption{NMSE versus received SNR ($P = 32$)}
        \label{fig:wb_snr}
    \end{subfigure}
    \caption{Comparison under the wideband model.}
    \label{fig:wb}
\end{figure*}

Table~\ref{tab:sim-params} summarizes the parameters. The TX positions are drawn from ten equal-width distance bins over $[18, 200)$~m, with 50 samples per bin, 
between the radiating near-field limit $0.62\sqrt{D_{\mathrm{a}}^3/\lambda_{\mathrm{c}}} \approx 18$~m of the array~\cite{balanis2016antenna}, with aperture $D_{\mathrm{a}} = N_{\mathrm{bs}} d$, and the farthest labeled vehicle, well within the Fraunhofer distance $2D_{\mathrm{a}}^2/\lambda_{\mathrm{c}} \approx 655$~m.
% between the radiating near-field limit $0.62\sqrt{D^3/\lambda_{\mathrm{c}}} \approx 18$~m of the array~\cite{balanis2016antenna} and the farthest labeled vehicle, well within the Fraunhofer distance $2D^2/\lambda_{\mathrm{c}} \approx 655$~m. 
For each TX position, $\sigma^2$ is set so that the average per-measurement received SNR at the combiner output equals the target value. Unless stated otherwise, $P = 32$ and the SNR is 10~dB.

Both narrowband and wideband variants of V-CL and V-NMR are evaluated, as the polar-domain benchmarks exist in both forms. The narrowband variants, labeled (NB), evaluate the channel at the carrier wavenumber $k_{\mathrm{c}}$ on every subcarrier and allow a like-for-like comparison with the narrowband benchmarks of~\cite{cui2022channel}. The wideband variants, labeled (WB), use the per-subcarrier $k_m$ of~\eqref{eq:channel} and show the gain of accounting for the frequency dependence of the channel. Accordingly, V-CL (NB) and V-NMR (NB) are compared against P-SOMP and P-SIGW~\cite{cui2022channel}, V-CL (WB) against BPD~\cite{cui2023near} and NBA-OMP~\cite{elbir2023nba}, and V-NMR (WB) against St-P-SIGW~\cite{chen2024segmented}, with matching on-grid or off-grid structure.
P-SOMP, P-SIGW, NBA-OMP, and St-P-SIGW share the polar grid of \cite{cui2022channel} (1,792 atoms) and BPD uses that of \cite{cui2023near} (3,584 atoms), both with the closest distance ring at about 18~m, and St-P-SIGW uses segments of 64 subcarriers, its best among 8 to 64. All benchmarks recover a single path, matching the channel. As a reference, Genie-LS knows the true TX position and estimates only the gains via~\eqref{eq:lsgain}. The vision-aided estimator of \cite{zhang2025vision} is not included, as it requires a neural network trained on labeled images, whereas the proposed methods require no training.

\subsection{Narrowband Model}
\label{sec:same}

In Fig.~2a, V-NMR (NB) matches Genie-LS (NB) at every pilot length and approaches the narrowband model floor at about $-22$~dB, whereas P-SOMP saturates at about $-5.5$~dB and P-SIGW approaches this floor as $P$ grows, remaining 1.3~dB above it at $P = 64$. The gain of V-NMR (NB) over P-SIGW is therefore largest when pilots are scarce, reaching 9.0~dB at $P = 8$. At $P = 8$, V-NMR (NB) is within 0.5~dB of P-SIGW at $P = 64$, i.e., with an eighth of the pilot overhead. Fig.~\ref{fig:nb_snr} shows the NMSE versus received SNR under the narrowband model. P-SOMP and P-SIGW saturate at about $-5.8$ and $-19.1$~dB, whereas V-NMR (NB) follows Genie-LS (NB) down to the model floor of $-22.6$~dB. Since V-NMR (NB) attains Genie-LS (NB), this floor is due to the narrowband model itself rather than to the vision-aided search, which motivates the wideband variants.

\subsection{Wideband Model}
\label{sec:wb}

In Fig.~\ref{fig:wb_pilot}, V-NMR (WB) coincides with Genie-LS (WB) and, free of the narrowband floor, decreases by about 3~dB per doubling of $P$, down to $-34.1$~dB at $P = 64$. 
At $P = 32$, the wideband variants improve on their narrowband counterparts by 3.3~dB for V-CL and 9.0~dB for V-NMR.
On-grid, BPD and NBA-OMP remain at $-6.2$ and $-5.8$~dB, respectively, against $-23.3$~dB for V-CL (WB); since NBA-OMP uses the same per-subcarrier atoms as V-CL (WB) and differs only in the candidate set, this 17.5~dB gap is due to the vision-aided grid. Off-grid, St-P-SIGW, which refines on the polar grid of P-SIGW, reaches $-15.0$~dB, 16.1~dB above V-NMR (WB).
 
Fig.~\ref{fig:wb_snr} shows the NMSE versus received SNR under the wideband model. BPD and NBA-OMP saturate at about $-6$~dB and St-P-SIGW at about $-15$~dB, whereas V-NMR (WB) follows Genie-LS (WB) over the entire range, decreasing by 10~dB per 10~dB of SNR down to $-51.1$~dB at 30~dB SNR, i.e., 36~dB below St-P-SIGW and 45~dB below BPD and NBA-OMP. Since the wideband model removes the floor of the narrowband one, the error of V-NMR (WB) is limited only by noise, so its advantage keeps growing with SNR.

\subsection{Grid Efficiency and Runtime}
\label{sec:grid}

\begin{table}[t]
\centering
\caption{Minimum Number of Grid Atoms to Reach Each Target NMSE ($P = 32$, SNR = 10~$\mathrm{dB}$)}
\label{tab:atoms-vs-nmse}
\setlength{\tabcolsep}{2pt}
\renewcommand{\arraystretch}{1.1}
\begin{tabular*}{\columnwidth}{@{\extracolsep{\fill}}cl cccccc@{}}
\toprule
& \multicolumn{1}{c}{\multirow{2}{*}{Method}} & \multicolumn{6}{c}{Target NMSE (dB)} \\
\cmidrule(l){3-8}
& & $-5$ & $-10$ & $-15$ & $-20$ & $-25$ & $-30$ \\
\midrule
\multirow{4}{*}{\rotatebox[origin=c]{90}{\footnotesize Narrowband}}
& \textbf{V-CL (NB)}$^\dagger$ & \textbf{9} & \textbf{16} & \textbf{36} & \textbf{144} & --- & --- \\
& P-SOMP~\cite{cui2022channel}$^\dagger$ & 1{,}792 & 7{,}147 & --- & --- & --- & --- \\
& \textbf{V-NMR (NB)}$^\ddagger$ & \textbf{4} & \textbf{4} & \textbf{9} & \textbf{9} & --- & --- \\
& P-SIGW~\cite{cui2022channel}$^\ddagger$ & 896 & 1{,}792 & 1{,}792 & --- & --- & --- \\
\midrule
\multirow{5}{*}{\rotatebox[origin=c]{90}{\footnotesize Wideband}}
& \textbf{V-CL (WB)}$^\dagger$ & \textbf{9} & \textbf{16} & \textbf{36} & \textbf{64} & \textbf{196} & --- \\
& BPD~\cite{cui2023near}$^\dagger$ & 3{,}584 & 7{,}154 & 28{,}574 & --- & --- & --- \\
& NBA-OMP~\cite{elbir2023nba}$^\dagger$ & 1{,}792 & 7{,}147 & --- & --- & --- & --- \\
& \textbf{V-NMR (WB)}$^\ddagger$ & \textbf{4} & \textbf{4} & \textbf{9} & \textbf{9} & \textbf{16} & \textbf{36} \\
& St-P-SIGW~\cite{chen2024segmented}$^\ddagger$ & 896 & 1{,}792 & 1{,}792 & --- & --- & --- \\
\bottomrule
\end{tabular*}
 
\vspace{2pt}
\begin{minipage}{\columnwidth}
\footnotesize
$^\dagger$On-grid. $^\ddagger$Off-grid. ---: not reached at any grid size tested.
\end{minipage}
\end{table}

\begin{figure}[t]
\centering
\includegraphics[width=\columnwidth]{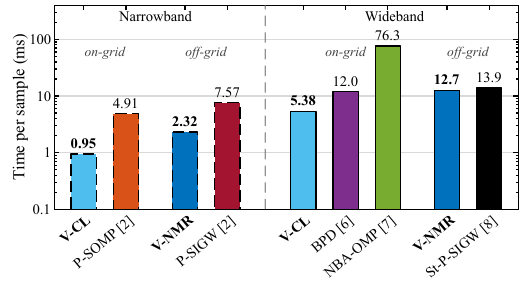}
\caption{Mean runtime per estimate ($P = 32$, received SNR $= 10$~dB), given the bounding box. V-NMR and P-SIGW include V-CL and P-SOMP; V-CL and V-NMR include Alg.~\ref{alg:vcl-grid}.}
\label{fig:timing}
\end{figure}
 
Table~\ref{tab:atoms-vs-nmse} reports the minimum number of grid atoms required to reach each NMSE, with the polar-domain grids evaluated over a range of angular resolutions. Under both models, the proposed algorithms need about two orders of magnitude fewer atoms than their benchmarks, and they reach $-20$~dB and beyond, which no benchmark attains at any grid size tested.

Fig.~\ref{fig:timing} reports the mean runtime per estimate in MATLAB on an Apple M4 CPU, with all methods vectorized to the same degree and the array-dependent dictionaries of \cite{cui2022channel, cui2023near, elbir2023nba} built offline, while their whitening, which depends on the combiner of each sample, is included in every runtime. The proposed methods run 2.2$\times$ to 14$\times$ faster than their benchmarks, except V-NMR (WB), which runs slightly faster than St-P-SIGW.
% V-CL (NB) and V-NMR (NB) run $5.2\times$ and $3.3\times$ faster than P-SOMP and P-SIGW, respectively. Under the wideband model, V-CL (WB) runs $2.2\times$ faster than BPD and $14\times$ faster than NBA-OMP, and V-NMR (WB) runs in 12.7~ms, against 13.9~ms for St-P-SIGW.

\section{Conclusion}
\label{sec:conclusion}

This paper presents a vision-aided framework for near-field channel estimation in wideband XL-MIMO systems, in which a camera image confines the candidate TX positions to a small grid on the object carrying the TX. On this grid, V-CL provides an on-grid estimate and V-NMR refines it off the grid. Under both the narrowband and wideband models, the proposed algorithms outperform polar-domain benchmarks in NMSE and pilot overhead, with V-NMR attaining the genie-aided reference, while requiring far fewer grid atoms at a lower or comparable runtime. Future work may address near-field TX identification, imperfect 3D detection and camera calibration, and multipath channels beyond the single LoS path considered here.

% This paper presents a vision-aided framework for near-field channel estimation in wideband XL-MIMO systems, in which a camera image confines the candidate TX positions to a small grid on the object carrying the TX. On this grid, V-CL provides an on-grid estimate and V-NMR refines it off the grid. Under the narrowband model, the proposed algorithms achieve up to 9~dB lower NMSE than the best polar-domain benchmarks and nearly match the best of them with an eighth of the pilot overhead. Under the wideband model, they outperform the wideband benchmarks by over 15~dB, with V-NMR attaining the genie-aided reference. In both cases, they require two orders of magnitude fewer grid atoms at a lower or comparable runtime. Future work may address near-field TX identification, imperfect 3D detection and camera calibration, and multipath channels beyond the single LoS path considered here.

\section*{Acknowledgment}

This work was supported by the Maria de Maeztu Units of Excellence Programme CEX2021-001195-M funded by MICIU/AEI/10.13039/501100011033, and by AGAUR.

% \angel{Use journalAbbreviations.bib for the standard abbreviations for journal and conference names}

% \textcolor{blue}{Todo: Remove some since I added new ref}
\bibliographystyle{IEEEtran}
\bibliography{References}

\end{document}